\documentclass{article}

\usepackage[preprint]{neurips_2020}
\usepackage[utf8]{inputenc}
\usepackage{amsmath}
\usepackage{amssymb}
\usepackage{algpseudocode}
\usepackage{algorithm}
\usepackage{hyperref}       
\usepackage{graphicx}
\usepackage[export]{adjustbox}
\usepackage{subcaption}

\setcitestyle{square,numbers,sort&compress}

\title{Robust softmax aggregation on blockchain based federated learning with convergence guarantee}

\author{%
  Huiyu Wu\\
  Department of Industrial Engineering and Management Sciences \\
  Northwestern University \\
  \texttt{huiyuwu2025@u.northwestern.edu} \\
  \And
  Diego Klabjan\\
  Department of Industrial Engineering and Management Sciences \\
  Northwestern University \\
  \texttt{d-klabjan@northwestern.edu} \\
}
\date{}
\newtheorem{assumption}{Assumption}
\newtheorem{theorem}{Theorem}
\newtheorem{proposition}{Proposition}
\newtheorem{lemma}{Lemma}
\newtheorem{corollary}{Corollary}

\begin{document}

\maketitle

\begin{abstract}
Blockchain based federated learning is a distributed learning scheme that allows model training without participants sharing their local data sets, where the blockchain components eliminate the need for a trusted central server compared to traditional Federated Learning algorithms. In this paper we propose a softmax aggregation blockchain based federated learning framework. First, we propose a new blockchain based federated learning architecture that utilizes the well-tested proof-of-stake consensus mechanism on an existing blockchain network to select validators and miners to aggregate the participants' updates and compute the blocks. Second, to ensure the robustness of the aggregation process, we design a novel softmax aggregation method based on approximated population loss values that relies on our specific blockchain architecture. Additionally, we show our softmax aggregation technique converges to the global minimum in the convex setting with non-restricting assumptions. Our comprehensive experiments show that our framework outperforms existing robust aggregation algorithms in various settings by large margins. 
\end{abstract}

\section{Introduction}
Federated learning (FL) is a machine learning schema that allows participants to collectively train a model using all available data without the need of actually sharing potentially sensitive data to each other \cite{FL}. The central server aggregates the participants' model updates without accessing their local data. This is particularly favorable as data privacy issues become major concerns in the modern world \cite{privacy}. It has a wide range of applications including predictive healthcare, learning user behavior on cell phones, and autonomous vehicles \cite{floverview}. However, traditional FL frameworks rely on a central server to coordinate the training process and suffer from having a single point of failure. If the server becomes compromised either by external attack or network failures, the entire FL process fails. To eliminate the need of trusted centralized servers, Blockchain based federated learning (BCFL) architectures are proposed \cite{bcflsurvey}.

Blockchain technology has been first used by Bitcoin and has since gained popularity in many fields such as cryptocurrency, healthcare, and finance \cite{bcflsurvey, bitcoin}. Blockchains are fully decentralized and can hold immutable records in blocks. Additionally, participants can earn rewards based on contribution. BCFL builds on the recent popularity and success of the blockchain technology and utilizes consensus mechanisms such as Proof-of-Work (PoW) or Proof-of-Stake (PoS) to synchronize model updates, therefore enabling participants to aggregate local models without the need of a trusted centralized entity. BCFL has the additional benefit of being resilient to a single point of failure, and encouraging participation to be trustworthy by providing incentives \cite{newbcflsurvey}. However, many BCFL schemes create their own rewards based on their blockchains which do not have much value out side of the scope. Additionally, BCFL in itself cannot defend against adversarial attacks, and recent works on robust aggregation methods as defending schemes are limited to specific settings and are not applicable to more general cases. 

To solve the above limitations of existing frameworks, we propose a softmax aggregation blockchain based federated learning framework named SABFL (Softmax Aggregation on Blockchain based Federated Learning). It relies on any existing blockchain network to handle the rewards, it utilizes the PoS consensus mechanism, and has a softmax robust aggregation method designed specifically for the architecture. Our blockchain structure has three types of participants: workers, validators, and miners. In a PoS scheme, validators and miners are generally viewed as trusted parties because they are at risk of losing their staking if they are dishonest. They are also participants with their own local data sets selected based on the size of their stakings. We further assume that participants with more significant stakes have larger and more representative training and validation data sets. Workers utilize only training data. Therefore, if we use the validators' data sets to approximate the population data set, and calculate the approximated population loss for each of the workers, the loss values can be used to describe the efforts of workers. Finally, we design a softmax based approach to map the approximated loss values to weights assigned to each worker's model update that are used in final aggregation.

In the experiments, we examine the performance of SABFL in a wide range of settings with participants having iid or non-iid local data sets using different machine learning models on various tasks. We show that SABFL achieves better performance by a large margin when compared to existing works. Crucially, we also explore the less studied areas including when up to half validators become dishonest and when clients have heterogeneity in data set sizes.

We summarize our main contributions as follows. First, we propose SABFL, a new blockchain based FL framework. It has all the benefits of BCFL architectures and the additional merit of providing universally accepted rewards. Second, we introduce a softmax based aggregation technique that is robust to data heterogeneity in both size and distribution, and performs well when there are malicious attacks. Additionally, we show that the softmax aggregation technique converges to the global minimum in the convex setting under mild assumptions. Third, our experiments expose that clients with varying sample sizes can seriously harm the accuracy of traditional FL schemes, which has been largely overlooked thus far. The experiment results also corroborate the claim that softmax aggregation is robust in different situations including even when up to half of the validators are colluding with malicious participants to carry out an attack.

This paper is organized as follows. Section \ref{lit_review} introduces the related works and how SABFL is different and better in comparison, Section \ref{operations} discusses the operations of SABFL including its underlying blockchain framework and the softmax aggregation technique; it also presents the convergence result. Section \ref{experiments} explains the comprehensive experiment design and presents the superior performance of SABFL, and finally Section \ref{conclusion} summarizes the key findings.

\section{Related Works}\label{lit_review}
Our work is related to BCFL architectures and the robustness of FL systems. Chen et al also focus on BCFL and robustness of its framework. They propose VBFL \cite{vbfl} which utilizes the differences of training accuracy between the workers and the validators to set a hard threshold that is then used to determine a binary vote of the workers' uploaded gradients. They also devise a clever reward mechanism that is based on the votes and the local training data set sizes of the participants. However, the hard threshold is fixed across the global rounds and is tested only on one data set with one model when in fact it can depend on the global round number and can vary significantly if the underlying data set or model changes. Additionally, sample size information can be sensitive in FL applications and cannot be easily verified.
Another limitation is that their participants have iid data sets, and when malicious participants are selected as validators, they act honorably in all but one experiment. We improve these drawbacks by introducing non-iid data sets, malicious validators in our testing, and an algorithm that captures such settings. Lastly, the gradient updates in VBFL are either included or excluded in the aggregation computation, and this may discourage participants with less computing resources and is not suitable for some BCFL applications.

Li et al propose BFLC \cite{bflc}, the concept of a new BCFL framework. They discuss utilizing a few honest participants' data as validation set to perform k-fold validation. However, k-fold validation is not resource efficient and it is uncertain how to select honest workers.  Other works on BCFL or decentralized FL are primarily focused on the blockchain architecture, or the operations and interactions of different participants \cite{blockFL, baffle, dfl, create, p2p, braintorrent, graph}. Although they may introduce participant reputation mechanisms to handle adversarial attacks, the experiments with malicious participants are often limited and are not the main focus. Other BCFL works design novel consensus mechanisms to ensure system security and robustness. For example, Kang et al \cite{pov} introduce proof of verification that can validate gradient updates using a global test set. However, this framework requires a global trusted authority, thus diminishing the decentralized nature of the BCFL schema. Additionally, the global test set is not always available in many BCFL applications. Similarly, Chen et al research the security of proof of elapsed time \cite{poet}. It also requires a trusted computing technology and the system can be jeopardized with a relatively small fraction of participants. 

Several papers suggest that poisoning and Byzantine attacks are serious threats to the FL paradigm, and this is also true in the BCFL framework \cite{vbfl, threats, flattacksurvey, analyzingattack}. Robust aggregation techniques can counter the effect of such attacks. Some popular methods include Krum \cite{krum}, trimmed mean \cite{median}, and median \cite{median}. These methods also provide convergence guarantee under mild assumptions. Other methods include Foolsgold \cite{foolsgold} and a more recent Mandera algorithm \cite{mandera}. However, we show that existing methods either are not suitable in the BCFL setting, or they perform worse when compared to SABFL in our comprehensive experiments. This is due to the robustness of SABFL against the number and roles of malicious participants as well as the heterogeneity in both the participant's data set sizes and distributions.

\section{SABFL Framework}\label{operations}
\subsection{Architecture}
Existing BCFL architectures can be grouped into three categories based on the roles of participants and blockchains \cite{bcflsurvey}: fully coupled, flexibly coupled, and loosely coupled BCFL schemes. In fully coupled BCFL, all nodes work in the blockchain network, in a flexibly coupled BCFL only miners participate in the generation of blocks, and in a loosely coupled BCFL, the blockchain is only used to distribute and record rewards. In all three architectures, rewards are generated when blocks are created similar to existing cryptocurrencies. However, one major drawback of these architectures is that the created rewards may not be widely accepted outside the BCFL scope thus disincentivizing participation. 

We propose SABFL which consists of two components. The first component is some existing blockchain network such as Ethereum that is used to handle validator and miner selection and reward management. Validators and miners are selected based on the stakes on the Ethereum network which are easily verifiable. Additionally, the rewards are given in Ethereum, which is universally accepted and has high liquidity thus encouraging participation. The nature of Ethereum also ensures that validators and miners are less likely to become malicious since they have more to lose when compared to traditional BCFL frameworks. The second component is the BCFL layer where the selected validators verify the participants' updates and upload them to miners to compile into blocks similar to the process of a fully coupled BCFL scheme. Therefore, SABFL maintains the decentralized nature avoiding single point of failure. 

In summary, the SABFL architecture has all the benefits of a decentralized FL schema. Additionally, the reliance on an existing blockchain network has the added benefits of encouraging participation with more enticing rewards, and ensuring security by disincentivizing malicious behavior of the validators and miners.


\subsection{SABFL}
In this section we discuss the components of SABFL in detail. If there are several miners, they do not impact SABFL and thus we assume that there is a single miner. The first step of SABFL is for an entity to initiate a machine learning task and post the total rewards on the existing blockchain network. Note that the task posting entity does not need to be trusted as long as the rewards are verified, thus the decentralized nature of BCFL is maintained. A block consists of validators' local loss values using the weights of the workers, the scores assigned to the weights of the workers by the miner, gradients from every worker, and the updated weight of the model for the next training round. A block can be validated by computing the scores assigned to the weights of the workers using the deterministic formula with the loss values of the validators, and by computing the aggregated gradient using our softmax aggregation formula.

After the task is established and rewards are verified, the training and block generation processes can begin which are summarized in Algorithm \ref{main_alg}. At the beginning of operations of block $r$, consider the set of all participants $P$. It is partitioned into the worker, validator, and miner set $P_w^r, P_v^r, P_m^r$, where $|P_m^r|=1$. The selection of validators and the miner is based on the stakes of the participants that are in the existing blockchain network, and is performed at the beginning of each block. The re-selection of validators and miners in each block prevents a `nondemocracy' side effect when certain clients always assume important roles \cite{democracy}. After the selection process, each worker $i$ can start its training with its local data set and sends the final local model weight $w_i^{r}$ to all of the validators. Validator $j$ evaluates the received weights using its own local data and sends the local loss values $\hat{F}_j(w_i^{r})$ from all the workers to the miner. Next, the miner aggregates the weights using softmax aggregation which is introduced in the next section. Finally, the miner compiles all the information received into a block and appends it to the blockchain and broadcasts the updated global model. This training block generation process continues until certain accuracy threshold or block number is reached. Finally, once the SABFL operations terminate, the training records can be used to distribute the rewards using predefined smart contracts on the existing blockchain network to ensure fairness and transparency \cite{blockflow}. 
\begin{algorithm}
\caption{SABFL}\label{main_alg}
\begin{algorithmic}
\State Initialize $\tilde{\mathbf{w}}^{0}$
\For{Block $r = 1, 2, 3, \dots$ }
    \State Randomly partition clients into $P_w^r, P_v^r, P_m^r$ based on staking
    \For{Worker $i = 1, 2, 3, \dots, K$} (Simultaneously)
        \State Set local model to $\tilde{\mathbf{w}}^{r-1}$ 
        \State Start local training process and obtain $\mathbf{w}^{r}_{i}$
        \State Broadcast $\mathbf{w}^{r}_{i}$ to all validators
    \EndFor
    \For{Validator $j = 1, 2, 3, \dots, V$}
        \For{Worker $i = 1, 2, 3, \dots, K$}
            \State Evaluate local loss value for worker $i$, $\hat{F}_j(\mathbf{w}^{r}_{i})$ 
            \State Send $\hat{F}_j(\mathbf{w}^{r}_{i})$ to the miner
        \EndFor
    \EndFor
    \State The miner updates $\tilde{\mathbf{w}}^{r}$ based on softmax aggregation, and creates and posts the new block
\EndFor
\end{algorithmic}
\end{algorithm}

\subsection{Softmax Aggregation}
The intuition behind softmax aggregation is that since the validators are also participants with own local data, we can approximate the population loss value using all of validators' local data sets. Then we can aggregate the workers' models using weighted average where weights are based on the approximated population loss.

According to the operations of SABFL, during the computation of each block the miner receives $V$ loss values $\{\hat{F}_j(w_i^r)\}_{j=1}^V$ for worker $i$'s model. The softmax aggregation starts with the miner computing $K$ values $\{\hat{F}(w_i^r)\}_{i=1}^K$ where $\hat{F}(w_i^r)=\frac{1}{V}\sum_{j=1}^V\hat{F}_j(w_i^r)$. Since validators are selected by the size of stakes, they generally should have larger stakes compared to other participants, and thus it is safe to assume participants with larger stakes are generally more influential entities and should have more representative data sets. With this assumption, $\hat{F}(w_i^r)$ should be a close approximation to the actual population loss of weights $w_i^r$ especially if $V$ is large. 

The softmax aggregation works by applying softmax function to the loss vector $(-\hat{F}(w_1^r), \dots, -\hat{F}(w_K^r))$ to obtain a probability vector $(\hat{p}_1^r, \dots, \hat{p}_K^r)$, namely
\begin{equation}
    \hat{p}_i^r=\frac{exp(-\hat{F}(\mathbf{w}^{r}_{i}))}{\sum_{s=1}^{K}exp(-\hat{F}(\mathbf{w}^{r}_{s}))}.
\end{equation}
Finally, we update the global model using the weighted average
\begin{equation}
    \tilde{\mathbf{w}}^{r+1} = \sum_{i=1}^{K} \hat{p}_i^r\mathbf{w}^{r}_{i}.
\end{equation}
We apply the softmax function to the negative loss values since we want the models with smaller loss values to have larger weights in the aggregation process, and the output vector of the softmax function can be conveniently used as weights in the final aggregation process. 

Since the weights in the softmax aggregation are posted to the blockchain, and larger weights generally reflect better models in our setting, the readily available vector $(\hat{p}_1^r, \dots, \hat{p}_K^r)$ is used to assign rewards for the work on block $r$. Additionally, validator approximated loss values $\{\hat{F}_j(w_i^r)\}_{j=1}^V$ are used to extract reward for validator $j$. Generally speaking, the smaller the average difference between $\{\hat{F}_j(w_i^r)\}_{i=1}^K$ and $\{\hat{F}(w_i^r)\}_{i=1}^K$, the larger the reward should be for validator $j$, although it can depend on data heterogeneity. The actual reward mechanism is beyond the scope of this discussion and should depend on the use case.

\subsection{Analysis} \label{analysis}
In this section we analyze the convergence property of softmax aggregation. We summarize the notations used in Table \ref{notations}. The goal of SABFL is to solve the optimization problem
\begin{equation}
    F^*=\underset{\mathbf{w} \in \mathbb{R}^d}{\text{min}} F(\mathbf{w}),
\end{equation}
where $F(\mathbf{w})=\mathbb{E}_{\xi \sim \mathcal{D}}f(\mathbf{w};\xi)$ is the population loss. We aim to establish the convergence of SABFL to $F^*$ in presence of convexity. We start by introducing our first assumption.
\begin{assumption} \label{validator}
Given weight $\mathbf{w}$ and any error $\epsilon>0$, any validator $j$ can compute an estimate of the population loss $\hat{F}_j(\mathbf{w})$ such that $|\hat{F}_j(\mathbf{w})-F(\mathbf{w})|<\epsilon$.
\end{assumption} 
\begin{table}[h!]
\centering
\begin{tabular}{ c|l } 
 \hline
 \multicolumn{1}{c|}{Symbols} & \multicolumn{1}{c}{Meaning} \\ 
 \hline 
 $B_r$ & Block $r$ local training batch size\\ 
 $\alpha_r$ & Block $r$ learning rate\\
 $f()$ & Loss function\\
 $\tilde{\mathbf{w}}^{r}$ & Random variable (RV) representing the global model weights at block $r$\\
 $\mathbf{w}_{i}^{r,t}$ & RV representing local weights at local step $t$ for participant $i$, at block $r$\\
 $\xi_{i, b}^{r, t}$ & RV representing local data for participant $i$, batch $b$, at block $r$ and local step $t$\\
 $\hat{p}_{i}^{r}$ & The score assigned to weights of participant $i$ at block $r$\\
 \hline
\end{tabular}
\caption{List of notations}
\label{notations}
\end{table}
The softmax aggregation technique depends on using validators' local data to approximate the population, and Assumption \ref{validator} ensures that the error of the approximation can be arbitrarily small. The validators are selected based on the stakes in the existing blockchain network, therefore, generally speaking the validators should be influential participants and should have larger and more representative data sets when compared to other participants. With these larger and more representative data sets the validators should be able to approximate the population loss with high precision.

With Assumption \ref{validator} in place, we can now present Algorithm \ref{modified_alg}, which is a more detailed version of SABFL that focuses on the components that are crucial for the convergence analysis. 
\begin{algorithm}
\caption{}\label{modified_alg}
\begin{algorithmic}
\State Initialize $\tilde{\mathbf{w}}^{0}$, select any $\epsilon, \tilde{\epsilon}>0$
\For{Block $r = 1, 2, 3, \dots$ }
    \For{Worker $i = 1, 2, 3, \dots, K$}
        \State Set $\mathbf{w}^{r,0}_{i} = \tilde{\mathbf{w}}^{r-1}$ 
        \For{Local round $t = 1, 2, 3, \dots, E$}
            \State Draw ${\xi}_{i}^{r, t}$, from distribution $\mathcal{D}$ and compute gradient estimator $\nabla f(\mathbf{w}^{r,t-1}_{i};{\xi}_{i}^{r, t})$
            \State Update $\mathbf{w}^{r,t}_{i} = \mathbf{w}^{r,t-1}_{i} - {\alpha_r} \nabla f(\mathbf{w}^{r,t-1}_{i};{\xi}_{i}^{r, t})$
        \EndFor
    \EndFor
    \For{Validator $j = 1, 2, 3, \dots, V$}
        \State Compute $\tilde{F}_j(\mathbf{w}^{r,E}_{i})$ such that $|\tilde{F}_j(\mathbf{w}_i^{r,E})-F(\mathbf{w}_i^{r,E})|<\tilde{\epsilon}$ for any $i$
        \State Compute $m_j^r=\max_i |\tilde{F}_j(\mathbf{w}^{r,E}_{i})| + \tilde{\epsilon}$
        \State Compute $\hat{F}_j(\mathbf{w}^{r,E}_{i})$ such that $|\hat{F}_j(\mathbf{w}_i^{r,E})-F(\mathbf{w}_i^{r,E})|<\frac{\epsilon}{2^rK^{5/2}m_j^r}$ for any $i$
    \EndFor
    \State The miner computes $\hat{F}(\mathbf{w}^{r,E}_{i}) = \frac{1}{V}\sum_{j=1}^{V}\hat{F}_j(\mathbf{w}^{r,E}_{i})$
    \State The miner computes $\hat{p}_i^r=\frac{exp(-\hat{F}(\mathbf{w}^{r,E}_{i}))}{\sum_{s=1}^{K}exp(-\hat{F}(\mathbf{w}^{r,E}_{s}))}$
    \State The miner updates $\tilde{\mathbf{w}}^{r} = \sum_{i=1}^{K} \hat{p}_i^r\mathbf{w}^{r,E}_{i}$
\EndFor
\end{algorithmic}
\end{algorithm}

The following assumptions are also needed in the convergence analysis and they are common in the literature of convergence analyses \cite{kavg, convergence, optimization}. Assumption \ref{lip} sets conditions on the population loss function, Assumption \ref{unbiased} ensures the stochastic gradients of the participants are unbiased, and finally, Assumption \ref{var_bound} places a bound on the variance of the stochastic gradients.

\begin{assumption} \label{lip}
The objective function $F\geq 0$ is strongly convex, continuously differentiable, and the gradient of $F$ is Lipschitz continuous with Lipschitz constant $L$.
\end{assumption}

\begin{assumption} \label{unbiased}
Stochastic gradient is an unbiased estimator for any fixed parameter $\mathbf{w}$ of the true gradient, namely we have 
$\mathbb{E}_{\xi_i^{r,t} \sim \mathcal{D}}\nabla f(\mathbf{w};\xi_i^{r,t})=\nabla F(\mathbf{w})$ for any $i, r, t$.
\end{assumption}

\begin{assumption} \label{var_bound}
The variance of the stochastic gradient norm is bounded by $M$, or 
$\mathbb{E}_{\xi_i^{r,t} \sim \mathcal{D}}||\nabla f(\mathbf{w};\xi_i^{r,t})||^{2} - ||\mathbb{E}_{\xi_i^{r,t} \sim \mathcal{D}}\nabla f(\mathbf{w};\xi_i^{r,t})||^{2}\leq M$ for any $i, r, t, \mathbf{w}$.
\end{assumption}

With the previous assumptions, we arrive at our main theorem. It shows that if we run Algorithm \ref{modified_alg}, the weighted average squared gradient norms go to $0$. We defer the proof of Theorem \ref{main_res} to Appendix.

\begin{theorem}\label{main_res}
Suppose Algorithm \ref{modified_alg} is run with Assumptions \ref{validator}-\ref{var_bound},, decreasing learning rates $\alpha_r$'s, and parameters satisfying 
\begin{equation*}
    \frac{L^2\alpha_r^2(E+1)(E-2)}{2}+L\alpha_r E\leq 1, \hspace{0.1cm}1-\delta \geq L^2\alpha_r^2 
\end{equation*}
for some constant $\delta \in (0,1)$. 
If $B_r=|\xi_i^{r,t}|$, then the weighted average squared gradient norms satisfy the following bound for all $R \in \mathbb{N}$
\begin{equation*}
\begin{split}
    \mathbb{E}\sum_{r=1}^{R}{\alpha_r}||\nabla  F(\Tilde{\textbf{w}}^{r-1})||^2\leq &\frac{2(F(\Tilde{\textbf{w}}_{0})-F^*)}{(E-1+\delta)}+\sum_{r=1}^{R}\frac{LE\alpha_r^2 M[6E+L(2E-1)(E-1)\alpha_r]}{6B_r(E-1+\delta)}\\
    &+\frac{2\epsilon}{(E-1+\delta)}.
\end{split}
\end{equation*}
Additionally, if 
\begin{equation*}
    \sum_{r=1}^{\infty}{\alpha_r}=\infty,  \sum_{r=1}^{\infty}\frac{\alpha_r^2}{B_r}\leq \infty, \sum_{r=1}^{\infty}\frac{\alpha_r^3}{B_r}\leq \infty,
\end{equation*}
then we have
\begin{equation*}
    \frac{1}{\sum_{s=1}^{R}\alpha_s}\mathbb{E}\sum_{r=1}^{R}{\alpha_r}||\nabla  F(\Tilde{\textbf{w}}^{r-1})||^2 \rightarrow 0,\,as\,R\rightarrow \infty.
\end{equation*}
\end{theorem}
Theorem \ref{main_res} shows the convergence of a weighted average of squared gradient norms and it cannot guarantee the convergence of the gradient norm itself. However, if we focus on the gradient norm at a randomly selected round we arrive at Corollary \ref{corollary}, showing the sampled gradient norms converge to $0$ in probability.
\begin{corollary}\label{corollary}
    Suppose the conditions of Theorem \ref{main_res} hold. Let $r(R) \in \{1,\dots, R\}$ represent a random index chosen with probabilities proportional to $\{\alpha_r\}_{r=1}^{R}$. Then $|\nabla  F(\Tilde{\textbf{w}}^{r(R)})| \rightarrow 0$ in probability as $R\rightarrow \infty$.
\end{corollary}

This Corollary follows from Theorem \ref{main_res} and Corollary 4.11 in \cite{optimization} and $E=1, B_r=1, \delta=0.01, \alpha_r=\frac{1}{\lceil 2L \rceil +r}$ is one example set of parameters that satisfy the requirements. To summarize, with convexity, conditions on validators' approximations, and common assumptions on the stochastic gradients, we are able to show that softmax aggregation techique converges to the global minimum. In the next section we examine some generalizations of SABFL.

\subsection{Generalization} \label{accuracy_extension}
It is possible to apply the softmax function to the approximated accuracy instead of the population loss. If we let $\hat{A}_j(w_i^r)$ denote the accuracy of model weights $w_i^r$ using validator $j$'s local data, and the miner computes $\hat{A}(w_i^r)=\frac{1}{V}\sum_{j=1}^V\hat{A}_j(w_i^r)$, then the softmax accuracy aggregated next global weight is $\tilde{\mathbf{w}}^{r+1} = \sum_{i=1}^{K} \hat{p}_i^{r,A}\mathbf{w}^{r}_{i}$, where $\hat{p}_{i}^{r,A}=\frac{exp(-\hat{A}(\mathbf{w}^{r}_{i}))}{\sum_{s=1}^{K}exp(-\hat{A}(\mathbf{w}^{r}_{s}))}$.

The softmax aggregation technique can also be generalized to the traditional FL setting. It is trivial if there exists a large globally available test set. Otherwise, the central server can designate a few clients as validators based on accumulated reputation or some other criteria and send them all workers' model updates. Then the validators can send back the approximated population loss values and the central server can utilize the loss values to complete the softmax aggregation process. However, it is worth noting that without the existing blockchain network, this procedure is at risk of having a larger number of malicious validators when compared to SABFL.



\section{Experiments}\label{experiments}
\subsection{Design of Experiments}
The main focus of the experiments is to study the robustness of the softmax aggregation method in various settings. The experimental settings are summarized in Table \ref{setup}. The first six experiments are on the MNIST data set with a varying number of benign and malicious participant combinations. Additionally, each of these experiments is run with two versions with participants having different levels of heterogeneity in data size and distribution with details in the next section. Column `Options' lists the number and type of experiment. LH cases correspond to low heterogeneity in data (heavily non-iid), HH is low heterogeneity (mild non-iid), and character `E' represents more local training epochs. In total there are 16 experiments. The remaining three experiments aim to show the performance of softmax aggregation on different data sets and models including CNN, RNN, and a feed-forward neural network, and they are run with participants having iid data. We opted for 8 malicious clients since the goal of these experiments is to show that the underlying model does not affect the relative performance of SABFL, and the results of the MNIST experiments show that SABFL is robust against different portions of malicious clients. 

\begin{table}[h!]
\centering
\caption{Experimental Setup} \label{setup}
\begin{tabular}{ |c|c|c|c|c|c| } 
 \hline
 Exp Num & Dataset & Model &  Num Participants & Num Malicious & Options  \\ \hline
 1 & MNIST & CNN & 20 & 0 & LH, HH\\ \hline
 2 & MNIST & CNN & 20 & 3 & LH, HH\\ \hline
 3 & MNIST & CNN & 20 & 8 & LH, HH, HHE\\ \hline
 4 & MNIST & CNN & 40 & 6 & LH, HH\\ \hline
 5 & MNIST & CNN & 40 & 16 & LH, HH\\ \hline
 6 & MNIST & CNN & 10 & 4 & LH, HH\\ \hline
 7 & FashionMNIST & CNN & 20 & 8 & LH\\ \hline
 8 & Cover type & MLP & 20 & 8 & LH\\ \hline
 9 & IMDB & RNN & 20 & 8 & LH\\ \hline
\end{tabular}
\end{table} 

To compare softmax aggregation to existing robust aggregation techniques, we include Krum and median aggregation methods as benchmarks\cite{krum, median}. We do not include trimmed-mean since it has similar performance compared to median \cite{median}. We also considered Bulyan, Foolsgold, and the latest Mandera \cite{bulyan, foolsgold, mandera}. Bulyan uses a concept similar to that of Krum, Foolsgold can only handle one kind of attack at a time and fails when only one attacker is present whereas SABFL does not rely on the number of malicious participants, and Mandera relies on the non-deterministic k-means algorithms for clustering and is not suitable in the BCFL setting. We also include the vanilla FedAvg for comparison \cite{FL}. Since the participants have different sample sizes which can be sensitive in many FL applications, we include a simple aggregation technique that uses simple averages of model updates without weighting by the sample sizes.

Robust aggregation techniques like our softmax aggregation are designed to defend against poisoning attacks. In our experiments, the malicious participants perform the label flipping attack due to its effectiveness and general applicability. Compared to the sign flipping and mean shift attacks, label flipping does not require having information on the honest workers' updates, and when compared to Gaussian noise attacks, it does not rely on an ad-hoc covariance matrix. Additionally, label flipping is one kind of a model poisoning attack in our setting as it affects the gradient updates, and it is considered more effective than data poisoning attacks \cite{threats}. We also allow malicious workers to be selected as validators, and when selected, they collude with the malicious workers and evaluate loss values based on the flipped labels. We require that the portion of malicious validators is no more than half, because for a PoS scheme it is more rational to uphold the integrity of the system when one controls more than half of the stakes. We do not use a real blockchain such as Ethereum but instead we implement Algorithm \ref{main_alg} because the goal is to study the performance and robustness of the softmax aggregation method.

\subsection{Implementation Details}
To allow participants to have data sets with varying sizes and distributions, we use the data partition scheme in RADFed \cite{noniid}. Specifically, hyper-parameter $\lambda$ from RADFed is set to $1$ version one of the experiments and $0.1$ for version two, and in this case version two has larger data heterogeneity. The local training epoch $E$ is set to $4$ in all experiments except version three of experiment 3 where it is set to $8$. The learning rate is fixed to $0.01$ in all experiments. 

In the implementation, to ensure a fair stopping criteria for all the aggregation methods tested, we devise a novel stopping rule. For every single run let us denote the test set accuracy after training for $i$ global rounds as $a_i$. Then for $i\geq30$ we consider the ratio of the minimum over the maximum accuracy over the last $30$ rounds. Specifically, we compute the sequence $k_i=\frac{\min S_i}{\max S_i}$ where $S_i$ denotes the set $\{a_j\}_{j=i-29}^{i}$. We stop when $k_i$ decreases for the first time, say at round $T$, and treat $\max a_i$ for $i\leq T$ as the final accuracy for the specific run. Note that $k_i$ decreases if the accuracy of the new round is the new minimum of the past $30$ rounds or if the last $30$ rounds' accuracy values are not the maximum. In either case, it is reasonable to assume we have already achieved the maximum accuracy.

For training of MNIST, we use a CNN with 32 and 64 filters with size 5 $\times$ 5 and 512 hidden units for the fully connected layer. For FashionMNIST, we use a similar structure but with two hidden layers of size 600 and 120. The feed forward neural network for the Forest Cover Type data set has two hidden layers of size 1,000 and 500. Finally, the recurrent neural network for IMDB sentiment classification consists of a 64 dimensional embedding layer, 2 layers of LSTM, and a hidden layer of size 256. All of the architectures selected are able to reach a final accuracy similar to the state-of-the-art in the traditional machine learning setting. Finally, the framework has been implemented in PyTorch using Google Colab with Nvidia Tesla T4 GPUs and Intel(R) Xeon(R) CPU @ 2.00GHz. 

\subsection{Results}
We run each experiment five times and report the average final accuracy. In this section we present our main experimental findings, and the complete results are included in the Appendix. The first finding is that clients having various samples sizes can be a significant threat to FL systems that has been overlooked in the past studies, and softmax aggregation is a defense. Although vanilla FL can be considered having the perfect defense against varying sample sizes, in real-world cases, samples sizes of FL participants can be sensitive business information and should not be shared. Existing works on aggregation methods are less studied on non-iid data, and even more so in cases where participants have different sample sizes. The study of the effect of size differences on FL systems is summarized in Table \ref{exp1_res}. The numbers in bold represent the best performance and those in italics the second best. It is expected that vanilla performs best since it is using private information, the number of local samples. In the first row, there is no malicious participant, but clients have varying data set sizes. Therefore, the accuracy differences between vanilla and simple aggregation can be considered as the impact of size differences, and the impact can be as high as $3.1\%$ in the final accuracy. We note that in the second row, vanilla aggregation achieves a $95.63\%$ final accuracy which is still higher than simple aggregation's performance in the first row. Since the only difference in the two settings is the additional 3 malicious participants, we conclude that the effect of having participants with varying sample sizes can be as harmful as having at least 3 malicious participants. From the table, we also notice that softmax aggregation is the only method that has final accuracy between vanilla and simple aggregation, indicating it is capable of handling clients with varying sample sizes. We also conclude that softmax aggregation performs well when no malicious clients are present.

\begin{table}[h!]
\centering
\caption{Selected results (\%)} \label{exp1_res}
\begin{tabular}{ |c|c|c|c|c|c| } 
 \hline 
 Option & Softmax &  Vanilla & Simple & Median & Krum  \\ \hline
 Exp1-LH & \emph{97.43} & \textbf{98.56} & 97.36 & 96.42 & 97.17 \\ \hline
 Exp1-HH & \emph{96.37} & \textbf{98.40} & 95.30 & 88.29 & 89.19\\ \hline
 Exp2-HH & \emph{95.16} & \textbf{95.63} & 94.75 & 85.93 & 88.72\\ \hline
\end{tabular}
\end{table} 

\begin{figure}[h]
\begin{subfigure}{0.5\textwidth}
\includegraphics[width=0.9\linewidth, height=4.5cm]{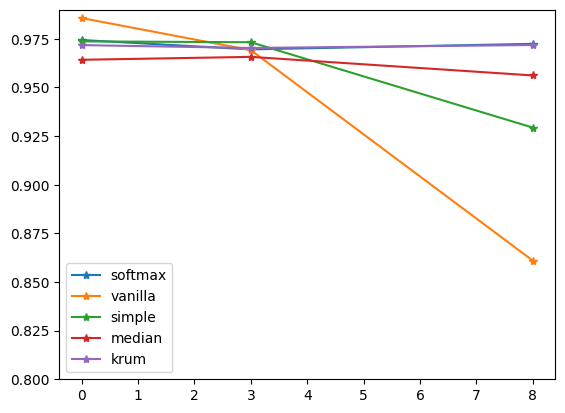} 
\caption{LH, $\lambda=1$}
\label{123results_a}
\end{subfigure}
\begin{subfigure}{0.5\textwidth}
\includegraphics[width=0.9\linewidth, height=4.5cm]{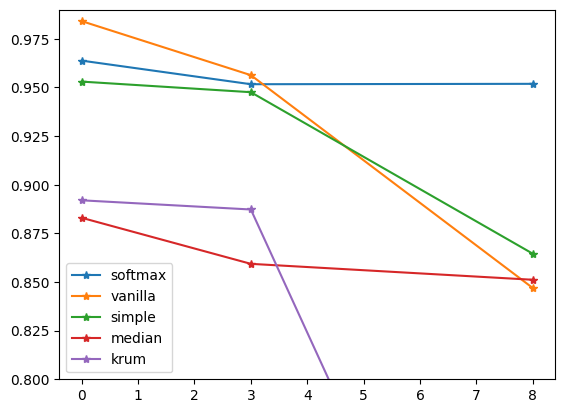}
\caption{HH, $\lambda=0.1$}
\end{subfigure}
\caption{Trends of performance with increasing number of malicious participants on MNIST.}
\label{123_results}
\end{figure}

We have also discovered that softmax aggregation is a great defense against malicious participants. Figure \ref{123_results} plots the trends of the final accuracy of different aggregation methods with increasing portions of malicious participants from $0\%$ to $40\%$. The x-axis shows the number of malicious clients. From the plots we learn that although softmax aggregation is not the best method when no malicious participants are present, it is the best method by a large margin when there are malicious clients, especially if clients have more data heterogeneity when $\lambda=0.1$. It is also worth noting that Krum has great performance in the case where $\lambda=1$ (in Figure \ref{123results_a} Krum and softmax overlap), but it is less robust when clients have less homogeneous data. Additionally, the plots show median aggregation is actually quite robust; the trends are almost flat similar to softmax aggregation. However, the drawback of the median aggregation method is its lower accuracy in comparison to softmax aggregation. 

\begin{figure}[h]
  \centering
  \includegraphics[height=4.8cm]{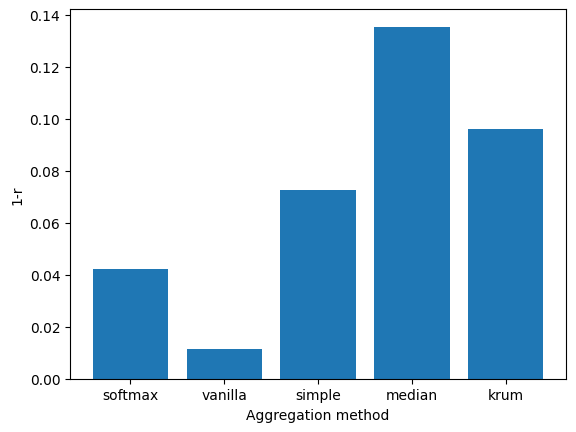}
  \caption{Robustness assessment}
  \label{robust_ratio}
\end{figure}

The third finding is that softmax aggregation is the only method that is robust against both data size and distribution heterogeneity simultaneously. For a given data set and number of malicious clients, the ratio of the final accuracy of LH over HH can be considered a measure of robustness since the only difference between the two versions is data heterogeneity in both distribution and size. We call this ratio $r$ the robustness score and ideally it should be as close to $1$ as possible. In Figure \ref{robust_ratio}, we plot the average $1-r$ score across all 16 experiments using different aggregation methods. We observe that our method has the second lowest value indicating a high robustness score next only to the vanilla FL aggregation, whereas median and Krum have low robustness scores. This is a crucial finding as many FL applications have participants with varying sample sizes and distributions, for example different autonomous cars may average varying miles and driving on different types of roads, and it seems existing methods do not perform well in such applications.

\begin{figure}[h]
\begin{subfigure}{0.5\textwidth}
\includegraphics[width=0.9\linewidth, height=4.5cm]{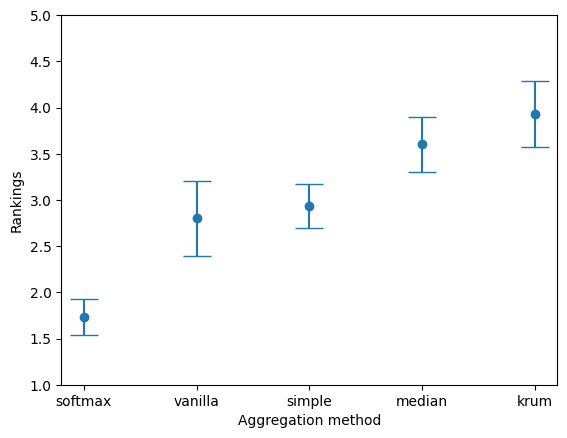} 
\caption{All experiments}
\label{all_rankings}
\end{subfigure}
\begin{subfigure}{0.5\textwidth}
\includegraphics[width=0.9\linewidth, height=4.5cm]{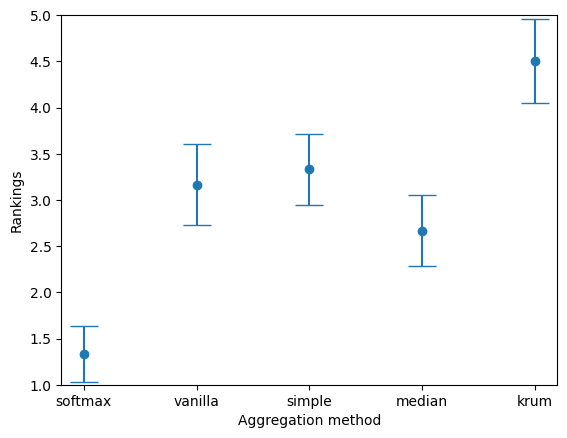}
\caption{With $40\%$ malicious}
\label{40rankings}
\end{subfigure}
\caption{Rankings statistics.}
\label{rankings}
\end{figure}

Finally, we find softmax aggregation has consistent relative performance across the various settings in our experiments. We rank the performance of each method from $1$ to $5$ across the $16$ different experimental settings, then we compute the average ranking statistics and present them in Figure \ref{rankings}. Figure \ref{all_rankings} contains the average rankings for all experiments, and Figure \ref{40rankings} only contains experiments where there are $40\%$ malicious participants, specifically experiments 3,5,7,8 and 9. We find that softmax aggregation not only has the best average ranking, it also has the smallest ranking deviation in both studies, indicating it has consistent superior performance in varying settings. We find that even when there are $40\%$ malicious clients, median aggregation performs relatively well in these previously untested situations. Krum suffers the most from robustness issues. Here although vanilla method has the highest robustness, its performance suffers in presence of malicious participants, and there are quite a few experiments with many malicious clients. 

We also conducted a study of the accuracy based softmax aggregation introduced in Section \ref{accuracy_extension} with the same setting as in Exp3-HH. We find accuracy based softmax aggregation achieves $94.52\%$ final accuracy, which is superior to the benchmark methods. We want to point out that softmax aggregation with accuracy loses the convergence property introduced previously.

Combining the findings from all the discussions, we conclude that the softmax aggregation method is the most robust in varying situations such as with and without malicious participants, clients having heterogeneous data sets in both size and distribution, and training with varying underlying models and tasks. Comparing Exp3-HHE and Exp3-HH results in Table \ref{accuracy_result}, we also conclude that the softmax aggregation method is robust against the number of local training epochs.

\section{Conclusion}\label{conclusion}
To conclude, in this paper we propose a new BCFL scheme and a softmax aggregation technique designed specifically for the proposed architecture. The reliance on existing blockchain networks ensures the participation rewards are universal and liquid. Softmax aggregation utilizes the architecture's approximated population loss to calculate the weights assigned to the workers' models. It does not assume honest validators and empirically we show that the aggregation technique is robust even when having up to half colluding malicious validators. Additionally, under mild assumptions we prove the convergence property of softmax aggregation. We also explore the less studied case when participants have samples of varying sizes. The results show it is a serious threat to existing FL schemes, specifically, the impact of varying data set sizes can be more severe than having $15\%$ malicious clients and existing robust aggregation methods do not perform well in these scenarios. We also research other robustness measures using different experimental settings and show that the softmax aggregation technique is by far the most robust, without compromising privacy and the decentralization assumptions of BCFL frameworks.


\begin{thebibliography}{32}
\providecommand{\natexlab}[1]{#1}
\providecommand{\url}[1]{\texttt{#1}}
\expandafter\ifx\csname urlstyle\endcsname\relax
  \providecommand{\doi}[1]{doi: #1}\else
  \providecommand{\doi}{doi: \begingroup \urlstyle{rm}\Url}\fi

\bibitem[Behera et~al.(2021)Behera, Upadhyay, Shetty, and den Otter]{p2p}
Monik~Raj Behera, Sudhir Upadhyay, Suresh Shetty, and Rob den Otter.
\newblock Federated learning using peer-to-peer network for decentralized
  orchestration of model weights.
\newblock \emph{TechRxiv}, 14267468, 2021.

\bibitem[Bhagoji et~al.(2019)Bhagoji, Chakraborty, Mittal, and
  Calo]{analyzingattack}
Arjun~Nitin Bhagoji, Supriyo Chakraborty, Prateek Mittal, and Seraphin Calo.
\newblock Analyzing federated learning through an adversarial lens.
\newblock In \emph{Proceedings of the 36th International Conference on Machine
  Learning}, volume~97, pages 634--643, 2019.

\bibitem[Blanchard et~al.(2017)Blanchard, El~Mhamdi, Guerraoui, and
  Stainer]{krum}
Peva Blanchard, El~Mahdi El~Mhamdi, Rachid Guerraoui, and Julien Stainer.
\newblock Machine learning with adversaries: Byzantine tolerant gradient
  descent.
\newblock In \emph{Advances in Neural Information Processing Systems},
  volume~30, 2017.

\bibitem[Bottou et~al.(2018)Bottou, Curtis, and Nocedal]{optimization}
L\'{e}on Bottou, Frank~E. Curtis, and Jorge Nocedal.
\newblock Optimization methods for large-scale machine learning.
\newblock \emph{SIAM Review}, 60\penalty0 (2):\penalty0 223--311, 2018.

\bibitem[Chen et~al.(2021)Chen, Asif, Park, Shen, and Bennis]{vbfl}
Hang Chen, Syed~Ali Asif, Jihong Park, Chien{-}Chung Shen, and Mehdi Bennis.
\newblock Robust blockchained federated learning with model validation and
  proof-of-stake inspired consensus.
\newblock \emph{arXiv}, 2101.03300, 2021.

\bibitem[Chen et~al.(2017)Chen, Xu, Shah, Gao, Lu, and Shi]{poet}
Lin Chen, Lei Xu, Nolan Shah, Zhimin Gao, Yang Lu, and Weidong Shi.
\newblock On security analysis of proof-of-elapsed-time (poet).
\newblock In \emph{Stabilization, Safety, and Security of Distributed Systems},
  pages 282--297, 2017.

\bibitem[Cui et~al.(2022)Cui, Su, Ming, Chen, Yang, Zhou, and Xiao]{create}
Laizhong Cui, Xiaoxin Su, Zhongxing Ming, Ziteng Chen, Shu Yang, Yipeng Zhou,
  and Wei Xiao.
\newblock Creat: Blockchain-assisted compression algorithm of federated
  learning for content caching in edge computing.
\newblock \emph{IEEE Internet of Things Journal}, 9\penalty0 (16):\penalty0
  14151--14161, 2022.

\bibitem[Fung et~al.(2020)Fung, Yoon, and Beschastnikh]{foolsgold}
Clement Fung, Chris J.~M. Yoon, and Ivan Beschastnikh.
\newblock Mitigating sybils in federated learning poisoning.
\newblock \emph{arXiv}, 1808.04866, 2020.

\bibitem[Gao and Pavel(2017)]{softmax_properties}
Bolin Gao and Lacra Pavel.
\newblock On the properties of the softmax function with application in game
  theory and reinforcement learning.
\newblock \emph{arXiv}, 1704.00805, 2017.

\bibitem[Kang et~al.(2020)Kang, Xiong, Jiang, Liu, Guo, Zhang, Niyato, Leung,
  and Miao]{pov}
Jiawen Kang, Zehui Xiong, Chunxiao Jiang, Yi~Liu, Song Guo, Yang Zhang, Dusit
  Niyato, Cyril Leung, and Chunyan Miao.
\newblock Scalable and communication-efficient decentralized federated edge
  learning with multi-blockchain framework.
\newblock In \emph{Blockchain and Trustworthy Systems}, pages 152--165, 2020.

\bibitem[Kim et~al.(2020)Kim, Park, Bennis, and Kim]{blockFL}
Hyesung Kim, Jihong Park, Mehdi Bennis, and Seong-Lyun Kim.
\newblock Blockchained on-device federated learning.
\newblock \emph{IEEE Communications Letters}, 24\penalty0 (6):\penalty0
  1279--1283, 2020.

\bibitem[Li et~al.(2022)Li, Han, Weng, Zheng, Li, Liu, Castiglione, and
  Li]{newbcflsurvey}
Dun Li, Dezhi Han, Tien-Hsiung Weng, Zibin Zheng, Hongzhi Li, Han Liu,
  Arcangelo Castiglione, and Kuan-Ching Li.
\newblock Blockchain for federated learning toward secure distributed machine
  learning systems: A systemic survey.
\newblock \emph{Soft Computing}, 26\penalty0 (9):\penalty0 4423–4440, 2022.

\bibitem[Li et~al.(2020{\natexlab{a}})Li, Sahu, Talwalkar, and
  Smith]{floverview}
Tian Li, Anit~Kumar Sahu, Ameet Talwalkar, and Virginia Smith.
\newblock Federated learning: Challenges, methods, and future directions.
\newblock \emph{IEEE Signal Processing Magazine}, 37\penalty0 (3):\penalty0
  50--60, 2020{\natexlab{a}}.

\bibitem[Li et~al.(2020{\natexlab{b}})Li, Huang, Yang, Wang, and
  Zhang]{convergence}
Xiang Li, Kaixuan Huang, Wenhao Yang, Shusen Wang, and Zhihua Zhang.
\newblock On the convergence of fedavg on non-iid data.
\newblock In \emph{The International Conference on Learning Representations},
  2020{\natexlab{b}}.

\bibitem[Li et~al.(2021)Li, Chen, Liu, Huang, Zheng, and Yan]{bflc}
Yuzheng Li, Chuan Chen, Nan Liu, Huawei Huang, Zibin Zheng, and Qiang Yan.
\newblock A blockchain-based decentralized federated learning framework with
  committee consensus.
\newblock \emph{IEEE Network}, 35\penalty0 (1):\penalty0 234--241, 2021.

\bibitem[Liu et~al.(2022)Liu, Xu, and Wang]{threats}
Peng Liu, Xiangru Xu, and Wen Wang.
\newblock Threats, attacks and defenses to federated learning: issues, taxonomy
  and perspectives.
\newblock \emph{Cybersecurity}, 5:\penalty0 1--19, 2022.

\bibitem[McMahan et~al.(2017)McMahan, Moore, Ramage, Hampson, and Arcas]{FL}
Brendan McMahan, Eider Moore, Daniel Ramage, Seth Hampson, and Blaise Aguera~y
  Arcas.
\newblock {Communication-Efficient Learning of Deep Networks from Decentralized
  Data}.
\newblock In \emph{Proceedings of the 20th International Conference on
  Artificial Intelligence and Statistics}, volume~54, pages 1273--1282, 2017.

\bibitem[Mhamdi et~al.(2018)Mhamdi, Guerraoui, and Rouault]{bulyan}
El~Mahdi~El Mhamdi, Rachid Guerraoui, and S{\'e}bastien Rouault.
\newblock The hidden vulnerability of distributed learning in byzantium.
\newblock In \emph{International Conference on Machine Learning}, 2018.

\bibitem[Mugunthan et~al.(2022)Mugunthan, Rahman, and Kagal]{blockflow}
Vaikkunth Mugunthan, Ravi Rahman, and Lalana Kagal.
\newblock Blockflow: Decentralized, privacy-preserving, and accountable
  federated machine learning.
\newblock In \emph{Blockchain and Applications}, pages 233--242, 2022.

\bibitem[Nakamoto(2008)]{bitcoin}
Satoshi Nakamoto.
\newblock Bitcoin: A peer-to-peer electronic cash system.
\newblock \emph{Bitcoin.org}, 2008.

\bibitem[Qu et~al.(2021)Qu, Wang, Hu, and Cheng]{democracy}
Xidi Qu, Shengling Wang, Qin Hu, and Xiuzhen Cheng.
\newblock Proof of federated learning: A novel energy-recycling consensus
  algorithm.
\newblock \emph{IEEE Transactions on Parallel and Distributed Systems},
  32\penalty0 (8):\penalty0 2074--2085, 2021.

\bibitem[Ramanan and Nakayama(2020)]{baffle}
P.~Ramanan and K.~Nakayama.
\newblock Baffle : Blockchain based aggregator free federated learning.
\newblock In \emph{2020 IEEE International Conference on Blockchain}, pages
  72--81, 2020.

\bibitem[Roy et~al.(2019)Roy, Siddiqui, Pölsterl, Navab, and
  Wachinger]{braintorrent}
Abhijit~Guha Roy, Shayan Siddiqui, Sebastian Pölsterl, Nassir Navab, and
  Christian Wachinger.
\newblock Braintorrent: A peer-to-peer environment for decentralized federated
  learning.
\newblock \emph{arXiv}, 1905.06731, 2019.

\bibitem[Sah and Singh(2022)]{privacy}
Mukund~Prasad Sah and Amritpal Singh.
\newblock Aggregation techniques in federated learning: Comprehensive survey,
  challenges and opportunities.
\newblock In \emph{2022 2nd International Conference on Advance Computing and
  Innovative Technologies in Engineering}, pages 1962--1967, 2022.

\bibitem[Sikandar et~al.(2023)Sikandar, Waheed, Tahir, Malik, and
  Rafique]{flattacksurvey}
Hira~Shahzadi Sikandar, Huda Waheed, Sibgha Tahir, Saif U.~R. Malik, and Waqas
  Rafique.
\newblock A detailed survey on federated learning attacks and defenses.
\newblock \emph{Electronics}, 12\penalty0 (2), 2023.

\bibitem[Tian et~al.(2023)Tian, Guo, Zhang, and Al-Ars]{dfl}
Yongding Tian, Zhuoran Guo, Jiaxuan Zhang, and Zaid Al-Ars.
\newblock Dfl: High-performance blockchain-based federated learning.
\newblock \emph{arXiv}, 2110.15457, 2023.

\bibitem[Wang et~al.(2022)Wang, Lalitha, Javidi, and Koushanfar]{graph}
Xinghan Wang, Anusha Lalitha, Tara Javidi, and Farinaz Koushanfar.
\newblock Peer-to-peer variational federated learning over arbitrary graphs.
\newblock \emph{IEEE Journal on Selected Areas in Information Theory},
  3\penalty0 (2):\penalty0 172--182, 2022.

\bibitem[Wang and Hu(2021)]{bcflsurvey}
Zhilin Wang and Qin Hu.
\newblock Blockchain-based federated learning: A comprehensive survey.
\newblock \emph{arXiv}, 2110.02182, 2021.

\bibitem[Xue et~al.(2022)Xue, Klabjan, and Luo]{noniid}
Ye~Xue, Diego Klabjan, and Yuan Luo.
\newblock Aggregation delayed federated learning.
\newblock In \emph{2022 IEEE International Conference on Big Data}, pages
  85--94, 2022.

\bibitem[Yin et~al.(2018)Yin, Chen, Kannan, and Bartlett]{median}
Dong Yin, Yudong Chen, Ramchandran Kannan, and Peter Bartlett.
\newblock {B}yzantine-robust distributed learning: Towards optimal statistical
  rates.
\newblock In \emph{Proceedings of the 35th International Conference on Machine
  Learning}, volume~80, pages 5650--5659, 2018.

\bibitem[Zhou and Cong(2018)]{kavg}
Fan Zhou and Guojing Cong.
\newblock On the convergence properties of a k-step averaging stochastic
  gradient descent algorithm for nonconvex optimization.
\newblock In \emph{Proceedings of the Twenty-Seventh International Joint
  Conference on Artificial Intelligence}, pages 3219--3227, 2018.

\bibitem[Zhu et~al.(2023)Zhu, Zhao, Luo, Liu, and Deng]{mandera}
Wanchuang Zhu, Benjamin Zi~Hao Zhao, Simon Luo, Tongliang Liu, and Ke~Deng.
\newblock Mandera: Malicious node detection in federated learning via ranking.
\newblock \emph{arXiv}, 2110.11736, 2023.

\end{thebibliography}

\newpage

\section*{Appendix}
\subsection*{A. Complete Results} \label{full_results_appendix}
Table \ref{accuracy_result} contains the results from all experiments and aggregation methods. The results are consistent with other results, specifically, the softmax aggregation method is the most robust in varying situations. It either is the best performer or has similar accuracy to the best aggregation method. It is also worth noting again that the vanilla aggregation method, which performs well in a few tasks, utilizes private sample size information that is beneficial to the performance, and ignoring the vanilla method, softmax aggregation is the best performer in 13 out of the 16 tasks.
\begin{table}[h!]
\centering
\caption{Final Accuracy (\%)} \label{accuracy_result}
\begin{tabular}{ |c|c|c|c|c|c| } 
 \hline 
 Experiment number & Softmax &  Vanilla & Simple & Median & Krum  \\ \hline
 Exp1-LH & 97.43 & \textbf{98.56} & 97.36 & 96.42 & 97.17 \\ \hline
 Exp1-HH & 96.37 & \textbf{98.40} & 95.30 & 88.29 & 89.19\\ \hline
 Exp2-LH & 96.95 & 96.91 & \textbf{97.32} & 96.57 & 97.03\\ \hline
 Exp2-HH & 95.16 & \textbf{95.63} & 94.75 & 85.93 & 88.72\\ \hline
 Exp3-LH & \textbf{97.24} & 86.10 & 92.93 & 95.61 & 97.18\\ \hline
 Exp3-HH & \textbf{95.18} & 84.69 & 86.45 & 85.11 & 57.11\\ \hline
 Exp3-HHE & \textbf{98.01} & 86.16 & 93.81 & 97.70 & 80.17\\ \hline 
 Exp4-LH & 92.95 & \textbf{96.65} & 93.53 & 90.77 & 78.02\\ \hline
 Exp4-HH & 83.21 & \textbf{92.73} & 83.10 & 67.51 & 64.74\\ \hline
 Exp5-LH & 89.87 & 82.29 & 92.02 & 88.90 & \textbf{94.68}\\ \hline
 Exp5-HH & 81.77 & \textbf{83.61} & 76.25 & 65.83 & 64.91\\ \hline
 Exp6-LH & \textbf{98.04} & 83.25 & 93.53 & 96.18 & 39.72\\ \hline
 Exp6-HH & \textbf{97.31} & 82.25 & 90.22 & 96.83 & 59.08\\ \hline
 Exp7-LH & \textbf{89.05} & 83.78 & 82.42 & 88.57 & 52.13\\ \hline
 Exp8-LH & \textbf{72.78} & 71.85 & 71.18 & 71.70 & 72.54\\ \hline
 Exp9-LH & \textbf{81.92} & 80.85 & 80.45 & 79.77 & 56.83\\ \hline
\end{tabular}
\end{table}

\subsection*{B. Proof of Theorem \ref{main_res}} \label{appendix}
To prove Theorem \ref{main_res}, we need to first introduce the following results. 
\begin{proposition}\label{softmax_mean}
    For non-negative real numbers $x_1, x_2, x_3, \dots, x_n$, it holds that 
    \begin{equation*}
        \frac{1}{\sum_{j=1}^{n}exp(x_j)}\sum_{i=1}^{n}{exp(x_i)}x_i \geq \frac{\sum_{i=1}^{n}x_i}{n}.
    \end{equation*}
    Additionally, the equality only holds when $x_1=x_2=x_3=\dots=x_n$.
\end{proposition}
\emph{Proof.} We prove by induction. For the base case of $n=1$ the statement simplifies to $\frac{exp(x_1)}{exp(x_1)}x_1 \geq x_1$, and it is trivially true. For the induction step, without loss of generality, let us assume $x_{k+1}\geq x_k \geq x_{k-1} \geq\dots\geq x_1$. To simplify the notation, let $S_{m}=\sum_{i=1}^{m} exp(x_i)$. We have the following
\begin{equation*}
\begin{split}
    \sum_{i=1}^{k+1}\frac{exp(x_i)}{S_{k+1}}x_i &= \frac{exp(x_{k+1})}{S_{k+1}}x_{k+1}+ \sum_{i=1}^{k}\frac{exp(x_i)}{S_{k+1}}x_i \\ 
    &= \frac{exp(x_{k+1})}{S_{k+1}}x_{k+1} + \frac{S_{k}}{S_{k+1}} \sum_{i=1}^{k}\frac{exp(x_i)}{S_{k}}x_i \\
    &\geq \frac{exp(x_{k+1})}{S_{k+1}}x_{k+1} + \frac{S_{k}}{S_{k+1}} \frac{\sum_{i=1}^{k}x_i}{k} \\
    &\geq \frac{1}{k+1}x_{k+1} + \frac{k}{k+1} \frac{\sum_{i=1}^{k}x_i}{k} \\
    &= \frac{\sum_{i=1}^{k+1}x_i}{k+1}.
\end{split}
\end{equation*}
The first inequality is by the induction hypothesis and the equality holds only when $x_1=x_2=x_3=\dots=x_k$. For the second inequality, note that we assume $x_{k+1}\geq x_k \geq x_{k-1} \geq\dots\geq x_1$, therefore, we have $\frac{exp(x_{k+1})}{S_{k+1}}\geq \frac{1}{k+1}$, $x_{k+1}\geq \frac{\sum_{i=1}^{k}x_i}{k}$, and the equality is true only when $x_{k+1}=x_i$, for every $i$. It also uses the fact that $f(s)=sx_{k+1}+(1-s)\frac{\sum_{i=1}^{k}x_i}{k}$ is an increasing function. \null\hfill $\blacksquare$\\
\smallskip\\
This proposition shows the softmax weighted average is no less than arithmetic average, an immediate corollary follows. 
\begin{corollary}\label{neg_softmax_mean}
    For non-negative real numbers $x_1, x_2, x_3, \dots, x_n$, we have 
    \begin{equation*}
        \frac{1}{\sum_{j=1}^{n}exp(-x_j)}\sum_{i=1}^{n}{exp(-x_i)}x_i \leq \frac{\sum_{i=1}^{n}x_i}{n}.
    \end{equation*}
    Additionally, the equality only holds when $x_1=x_2=x_3=\dots=x_n$.
\end{corollary}
The proof follows by applying Lemma \ref{softmax_mean} with $-x_1, -x_2, -x_3, \dots, -x_n$.
\smallskip \\
We have to introduce another lemma to prove our main result. It is a special case of a result from the proof of Theorem 1 in \cite{kavg} when there is only one participant.
\begin{lemma}\label{kavg_res}
    \emph{(Result from the proof of Theorem 1 in \cite{kavg})} Suppose Assumptions \ref{lip}-\ref{var_bound} hold. For fixed $\mathbf{w}^0$, let $\Tilde{\mathbf{w}}= \mathbf{w}^0 - \alpha\sum_{t=1}^{E}\sum_{b=1}^{B} \nabla f(\mathbf{w}^{t-1};\xi_{i}^{r,t})$, and let $\alpha$ be such that 
    \begin{equation*}
        \frac{L^2\alpha^2(E+1)(E-2)}{2}+L\alpha E\leq 1, and\hspace{0.1cm}1-\delta \geq L^2\alpha^2 
    \end{equation*}
    for some constant $\delta \in (0,1)$. If $B=|\xi_{i}^{r,t}|$, then we have 
    \begin{equation*}
     \mathbb{E}F(\Tilde{\textbf{w}}) - F(\mathbf{w}^0)\leq-\frac{(E-1+\delta)\alpha}{2}||\nabla  F(\textbf{w}^0)||^2+\frac{L\alpha^2EM}{2B}(E+\frac{L(2E-1)(E-1)\alpha}{6}).
    \end{equation*}
\end{lemma}

With the previous results, we can now start the proof of the main result. \\

\emph{Proof of Theorem \ref{main_res}.} Consider the random variables of two consecutive rounds' global weights, $\Tilde{\textbf{w}}^{r}$ and $\Tilde{\textbf{w}}^{r-1}$, and let us define $p_i^r =\frac{exp(-F(\mathbf{w}^{r,E}_{i}))}{\sum_{s=1}^{K}exp(-F(\mathbf{w}^{r,E}_{s}))}$. Then we can bound the global objective by 
\begin{equation}\label{round decrease}
\begin{split}
    F(\Tilde{\textbf{w}}^{r}) - F(\Tilde{\textbf{w}}^{r-1}) &= F(\sum_{i=1}^{K} \hat{p}_i^r\mathbf{w}^{r,E}_{i}) - F(\Tilde{\textbf{w}}^{r-1})\\
    & \leq \sum_{i=1}^{K} \hat{p}_i^rF(\mathbf{w}^{r,E}_{i})- F(\Tilde{\textbf{w}}^{r-1}) \\
    & = \sum_{i=1}^{K} (p_i^r+(\hat{p}_i^r-p_i^r))F(\mathbf{w}^{r,E}_{i})- F(\Tilde{\textbf{w}}^{r-1}) \\
    & = \sum_{i=1}^{K} p_i^rF(\mathbf{w}^{r,E}_{i})- F(\Tilde{\textbf{w}}^{r-1}) + \sum_{i=1}^{K}(\hat{p}_i^r-p_i^r)F(\mathbf{w}^{r,E}_{i}).
\end{split}
\end{equation}
Since the softmax function is $1$-Lipschitz according to Proposition 4 in \cite{softmax_properties} and if we let $m_r=\min_j m_j^r$, then Algorithm \ref{modified_alg} implies $|F(\mathbf{w}^{r,E}_{i})|\leq m_r$ and $|\hat{F}(\mathbf{w}_i^{r,E})-F(\mathbf{w}_i^{r,E})|<\frac{\epsilon}{2^rK^{5/2}m_r}$, and then we have 
\begin{equation}\label{epi_bound}
\begin{split}
    \sum_{i=1}^{K}(\hat{p}_i^r-p_i^r)F(\mathbf{w}^{r,E}_{i}) & \leq \sum_{i=1}^{K}|\hat{p}_i^r-p_i^r||F(\mathbf{w}^{r,E}_{i})| \leq \sum_{i=1}^{K}K^{1/2}||\mathbf{\hat{p}}^r-\mathbf{p}^r||_2 |F(\mathbf{w}^{r,E}_{i})|\\
    & \leq \sum_{i=1}^{K}K^{1/2} ||\mathbf{\hat{F}}(\mathbf{w}^{r,E}) - \mathbf{F}(\mathbf{w}^{r,E})||_2 |F(\mathbf{w}^{r,E}_{i})|\\
    & \leq \sum_{i=1}^{K}K^{1/2} ||\mathbf{\hat{F}}(\mathbf{w}^{r,E}) - \mathbf{F}(\mathbf{w}^{r,E})||_1 |F(\mathbf{w}^{r,E}_{i})|\\ 
    & \leq \sum_{i=1}^{K}K^{1/2} K \frac{\epsilon}{2^rK^{5/2}m_r}m_r \\
    & = \frac{\epsilon}{2^r}.
\end{split}
\end{equation}
The bolded letters $\mathbf{\hat{p}}^r, \mathbf{\hat{p}}^r, \mathbf{\hat{F}}(\mathbf{w}^{r,E}), \mathbf{F}(\mathbf{w}^{r,E})$ denote the $K$ dimensional vectors. Next we apply Corollary \ref{neg_softmax_mean} to $F(\mathbf{w}^{r,E}_{1}), F(\mathbf{w}^{r,E}_{2}), F(\mathbf{w}^{r,E}_{3}), \dots, F(\mathbf{w}^{r,E}_{K})$, plug inequality (\ref{epi_bound}) into inequality (\ref{round decrease}), and obtain
\begin{equation*}\label{norm_bound}
\begin{split}
    F(\Tilde{\textbf{w}}^{r}) - F(\Tilde{\textbf{w}}^{r-1}) & \leq \frac{1}{K}\sum_{i=1}^{K} F(\mathbf{w}^{r,E}_{i}) - F(\Tilde{\textbf{w}}^{r-1}) +\frac{\epsilon}{2^r}\\
    & = \frac{1}{K} \sum_{i=1}^{K}(F(\mathbf{w}^{r,E}_{i}) - F(\Tilde{\textbf{w}}_{r^1})) + \frac{\epsilon}{2^r}.
\end{split}
\end{equation*}
We are interested in expected objective decrease, where the expectation is with respect to the set of random variables $\Xi_r = \{\xi_{i}^{r,t}\}_{(i,t)}$. We use the law of total expectation with respect to $r$. Participant $j$'s update is computed as 
\begin{equation*}
    \mathbf{w}^{r,E}_{j}= \Tilde{\textbf{w}}^{r-1} - \alpha_r\sum_{t=1}^{E} \nabla f(\mathbf{w}^{r,t-1}_{j};\xi_{j}^{r, t}).
\end{equation*}
We have 
\begin{equation}
\begin{split}
    \mathbb{E}_{\Xi_{r}}[F(\Tilde{\textbf{w}}^{r}) - F(\Tilde{\textbf{w}}^{r-1})|{\Xi_{r-1}}] 
    &\leq \frac{1}{K} \sum_{i=1}^{K} \mathbb{E}_{\Xi_{r}}[F(\mathbf{w}^{r,E}_{i}) - F(\Tilde{\textbf{w}}^{r-1})|{\Xi_{r-1}}]+\frac{\epsilon}{2^r}.
\end{split}
\end{equation}
Now we apply Lemma \ref{kavg_res} and the expectation of all random variables on both sides to obtain 
\begin{equation*}\label{eq11}
\begin{split}
         \mathbb{E}_{\Xi_{r}}[F(\Tilde{\textbf{w}}^{r}) - F(\Tilde{\textbf{w}}^{r-1})]&=\mathbb{E}_{\Xi_{r-1}}\mathbb{E}_{\Xi_{r}}[F(\Tilde{\textbf{w}}^{r}) - F(\Tilde{\textbf{w}}^{r-1})|{\Xi_{r-1}}]\\
         &\leq-\frac{(E-1+\delta)\alpha_r}{2}\mathbb{E}_{\Xi_{r-1}}||\nabla  F(\Tilde{\textbf{w}}^{r-1})||^2\\
         &\hspace{0.5cm} +\frac{L\alpha_r^2EM}{2B_r}(E+\frac{L(2E-1)(E-1)\alpha_r}{6}) + \frac{\epsilon}{2^r}.
\end{split}
\end{equation*}
Combining with the fact that $F^*-F(\Tilde{\textbf{w}}_{0})\leq F(\Tilde{\textbf{w}}_{r})-F(\Tilde{\textbf{w}}_{0})$, we obtain the following bound
\begin{equation*}
\begin{split} 
    \mathbb{E}\sum_{r=1}^{R}{\alpha_r}||\nabla  F(\Tilde{\textbf{w}}^{r-1})||^2\leq &\frac{2(F(\Tilde{\textbf{w}}_{0})-F^*)}{(E-1+\delta)}+\sum_{r=1}^{R}\frac{LE\alpha_r^2 M[6E+L(2E-1)(E-1)\alpha_r]}{6B_r(E-1+\delta)}\\
    &+\frac{2\epsilon}{(E-1+\delta)},
\end{split}
\end{equation*}
which completes the proof.  \hfill $\blacksquare$ \\

\end{document}